\global\def\draftcontrol{0}

   \def\versionno{  } 

\catcode`\@=11 

\expandafter\ifx\csname draftcontrol\endcsname\relax\global\def\draftcontrol{0} 
\fi 

{\count255=\time\divide\count255 by 60 
\xdef\hourmin{\number\count255} 
\multiply\count255 by-60\advance\count255 by\time 
\xdef\hourmin{\hourmin:\ifnum\count255<10 0\fi\the\count255}} 
\def\draftdate{\number\month/\number\day/\number\year\ \ \ \hourmin } 

\newcommand\makepapertitle{\par

  \begingroup 
    \renewcommand\thefootnote{\@fnsymbol\c@footnote}%
    \def\@makefnmark{\rlap{\@textsuperscript{\normalfont\@thefnmark}}}%
    \long\def\@makefntext##1{\parindent 1em\noindent 
            \hb@xt@1.8em{%
                \hss\@textsuperscript{\normalfont\@thefnmark}}##1}%
     \newpage 
     \global\@topnum\z@   
     \@makepapertitle 
     \thispagestyle{empty}\@thanks 
  \endgroup 
  \setcounter{footnote}{0}%
  \global\let\thanks\relax 
  \global\let\makepapertitle\relax 
  \global\let\@makepapertitle\relax 
  \global\let\@thanks\@empty 
  \global\let\@author\@empty 
  \global\let\@date\@empty 
  \global\let\@title\@empty 
  \global\let\title\relax 
  \global\let\author\relax 
  \global\let\date\relax 
  \global\let\and\relax 
  \def\version{\let\version\@version\@gobble} 
} 
\def\@makepapertitle{%
  \newpage 
   \ifnum\draftcontrol=1 {} 
   \version\versionno 
   \vskip 5em%
   \else 
   \hfill\hbox to 3.5cm {\parbox{5cm}{\@pubnum}\hss}%
   \vskip 3em%
   \fi 
   \begin{center}%
   \let \footnote \thanks 
      {\hskip -0\textwidth \hbox to 1\textwidth%
        {\centerline{\Large\bf{\noindent\@title}}}}%
     \vskip 2em%
     {\normalsize
       \lineskip .5em%
       \begin{tabular}[t]{c}%
         \@author 
       \end{tabular}\par}%
     \vskip 1em%
     {\@bstract}%
     \end{center}%
     \vfill
     \@date%
     \vskip 1.5em%
   \par 
} 

\gdef\@pubnum{} 
\def\pubnum#1{%
  \gdef\@pubnum{#1}} 

\gdef\@bstract{} 
\def\Abstract#1{%
  \gdef\@bstract{%
   \parbox{\textwidth-0pc}{%
   \centerline{\bf Abstract}\penalty1000 
   \noindent
   \renewcommand\baselinestretch{1.0} 
   {#1}}} 
} 

\gdef\@email{}
\def\email#1{%
   \gdef\@email{%
   Email: {\tt #1}}
}

\def\ps@paper{\let\@mkboth\@gobbletwo%
     \ifnum\draftcontrol=1 
        \def\@oddfoot{\hbox to \textwidth{\tiny \versionno \hfil\tiny\draftdate}%
        \hskip -\textwidth \hbox to \textwidth{\hfil\rm\thepage\hfil}}%
     \else\def\@oddfoot{\hbox to \textwidth{\hfil\rm\thepage\hfil}} 
     \fi 
     \let\@evenfoot\@oddfoot 
} 

\def\body{\clearpage 
          \pagestyle{paper} 
        } 
\newenvironment{acknowledgments}{%
\vskip 3.25ex 
\noindent {\bf Acknowledgments} 
} 

\def\@version#1{\ifnum\draftcontrol=1 
\typeout{}\typeout{#1}\typeout{} 
\vskip3mm\centerline{\hbox{\fbox{\normalsize{\tt DRAFT -- #1 -- } 
                   {\draftdate}}}}\vskip3mm 
\fi} 
\let\version\@version 
\long\def\eqlabel#1{\ifnum\draftcontrol=1 
                    \tag@false  
                    \tag*{(\theequation) \hbox to -0.2cm{\hspace{0cm}\small{#1}\hss}} 
                    \refstepcounter{equation}  
                    \edef\@currentlabel{\theequation} 
                    \ltx@label{#1}          
                    \else 
                    \label{#1} 
                    \fi 
                    } 
\let\st@bibitem\@bibitem 
\let\st@lbibitem\@lbibitem 
\ifnum\draftcontrol=1 
  \def\@bibitem#1{%
    \st@bibitem{#1}\a@@label{#1}\ignorespaces} 
  \def\@lbibitem[#1]#2{%
    \st@lbibitem[#1]{#2}\a@@label{#2}\ignorespaces} 
  \def\a@@label#1{%
    \gdef\a@lab{\smash{\normalfont\small#1}} 
    \ifvmode 
      \if@inlabel 
        \global\setbox\@labels\hbox{%
          \llap{\a@lab\let\a@lab\relax 
                \kern\@totalleftmargin\kern\marginparsep}%
          \box\@labels}%
      \fi 
    \fi} 
\fi 

\documentclass[12pt,letterpaper]{article} 

\usepackage{amsmath,bm,amsfonts,amssymb,array,calc,amsthm,rotating,cite}
\usepackage{graphicx} 

\renewcommand\baselinestretch{1.25} 
\renewcommand\section{\@startsection {section}{1}{\z@}%
                                   {-3.5ex \@plus -1ex \@minus -.2ex}%
                                   {2.3ex \@plus.2ex}%
                                   {\normalfont\large\bfseries}} 
\renewcommand\subsection{\@startsection{subsection}{2}{\z@}%
                                   {-3.25ex\@plus -1ex \@minus -.2ex}%
                                   {1.5ex \@plus .2ex}%
                                   {\normalfont\normalsize\bfseries}} 
\renewcommand\subsubsection{\@startsection{subsubsection}{3}{\z@}%
                                   {-3.25ex\@plus -1ex \@minus -.2ex}%
                                   {1.5ex \@plus .2ex}%
                                   {\normalfont\normalsize\it}} 
\renewcommand\paragraph{\@startsection{paragraph}{4}{\z@}%
                                   {-3.25ex\@plus -1ex \@minus -.2ex}%
                                   {1.5ex \@plus .2ex}%
                                   {\normalfont\normalsize\bf}} 
\renewcommand\subparagraph{\@startsection{subparagraph}{5}{\z@}%
                                   {-1.25ex\@plus -1ex \@minus -.2ex}%
                                   {0ex \@plus .2ex}%
                                   {\normalfont\normalsize\it}} 

\long\def\@makecaption#1#2{%
  \vskip\abovecaptionskip
  \sbox\@tempboxa{{\bf #1:} #2}%
  \ifdim \wd\@tempboxa >\hsize
    {\small\bf #1:} {\small #2}\par
  \else
    \global \@minipagefalse
    \hb@xt@\hsize{\hfil\box\@tempboxa\hfil}%
  \fi
  \vskip\belowcaptionskip}

\def\ie{{\it i.e.}} 
\def\eg{{\it e.g.}}

\def\ee{{\rm e}}
\def\ii{{\rm i}}
\def\complex{{\mathbb C}}

\newcommand\topa[2]{\genfrac{}{}{0pt}{0}{\textstyle #1}{\textstyle #2}} 

\def\revise#1       {\raisebox{-0em}{\rule{3pt}{1em}}%
                     \marginpar{\raisebox{.5em}{\vrule width3pt\ 
                     \vrule width0pt height 0pt depth0.5em 
                     \hbox to 0cm{\hspace{0cm}{%
                     \parbox[t]{4em}{\raggedright\footnotesize{#1}}}\hss}}}}

\newcommand{\beq}{\begin{equation}}
\newcommand{\eeq}{\end{equation}}
\newcommand{\bea}{\end{eqnarray}}
\newcommand{\eea}{\end{eqnarray}}

\newcommand{\beql}[1]{\begin{eqnarray}\label{#1}}

\newcommand{\eeql}{\end{eqnarray}}

\long\def\begdel#1\enddel{}

\def\del{\partial}

\def\cQ{{\mathcal Q}}
\def\cO{{\mathcal O}}
\def\cT{{\mathcal T}}
\def\cL{{\mathcal L}}
\def\cS{{\mathcal S}}

\def\al{\alpha}
\def\bfone{{\bf 1}}

\def\IP{{\mathbb P}}

\def\al{\alpha}
\def\be{\beta}

\def\weff{{\mathcal W}_{\rm eff}}
\def\wlg{W}
\def\str{{{\rm str}}}
\def\del{\partial}

\def\ZZ{{\mathbb Z}}
\def\zet{\ZZ}
\def\alb#1{^{(#1)}}
\def\href#1#2{#2}
\def\alf#1#2{{\alpha_{#1}^{#2}}}
\def\zta#1#2{{\zeta_{#1}^{#2}}}
\def\bpsi{{a}}
\def\la{\lambda}
\def\tht{\Theta}

\catcode`\@=12 
\begin{document}        
\pubnum{
  CERN--PH--TH/2004-162\\
  KCL-MTH-04-12\\
  NSF-KITP-04-106\\
  {\tt hep-th/0408243}\\
}
\title{
\parbox{\textwidth}{\centering Matrix Factorizations And Mirror Symmetry:\\
\centering The Cubic Curve}}
\author{
Ilka Brunner${}^{a,b}$, Manfred Herbst${}^a$, Wolfgang Lerche${}^a$, Johannes Walcher${}^c$
\footnote{After September 1: Institute for
Advanced Study, Princeton, New Jersey, USA}
\\[.4cm]
\it ${}^{a}$ Department of Physics, Theory Division\\ 
\it CERN, Geneva, Switzerland\\[0.2cm]
\it ${}^b$ Department of Mathematics\\
\it King's College, London, United Kingdom\\[0.2cm]
\it ${}^c$Kavli Institute for Theoretical Physics \\
\it University of California, Santa Barbara, California, USA \\[0.4cm] 
}

\Abstract{
We revisit open string mirror symmetry for the elliptic curve, using
matrix factorizations for describing $D$-branes on the $B$-model
side. We show how flat coordinates can be intrinsically defined in
the Landau-Ginzburg model, and derive the $A$-model partition
function counting disk instantons that stretch between three
$D$-branes. In mathematical terms, this amounts to computing the
simplest Fukaya product $m_2$ from the LG mirror theory.
In physics terms, this gives a systematic method for determining 
non-perturbative Yukawa couplings for intersecting brane configurations.
}

\date{August 2004}


\makepapertitle 

\body 

\version\versionno 


\section{Introduction}

The considerable recent progress in computing non-perturbative
superpotentials (and other holomorphic quantities) in $N=1$ string
vacua, has left behind several open questions concerning the general
systematics of open topological strings \cite{mirbook}.  The list includes, on the
technical side, the proper inclusion of boundary changing sectors,
associated with worldsheets spanning between different branes.  On
the conceptual side, it remains an outstanding question how to find,
in general, proper ''special coordinates'' of mirror symmetry on
the combined open-closed string parameter space. This latter problem
is severe not only if boundary changing sectors are included, but
even more so when deformations are obstructed and the notion of
flatness becomes an off-shell or merely infinitesimal question.
 
Recently, a promising approach for describing topological $D$-branes
in the $B$-model has been developped which is based on boundary
Landau-Ginzburg theory \cite{Kap,BHLS,Kap2,Kap3,dAAD,dAADF,HW,HLLmm,Herbst:2004ax,KapRo}, 
building on previous work \cite{Warner:1995ay,Govindarajan:1999js,Govindarajan:2000my,
Govindarajan:2000ef,Hori:2000ic} and \cite{kontsevich2,orlov}. 
It seems to capture all the relevant information about the 
category of topological $D$-branes of $B$-type, and has  
been successfully applied in particular to the topological minimal 
models, for which the complete effective superpotential on the disk 
has been determined \cite{HLLmm}. This was achieved by solving the open 
string version \cite{HLL} of the WDVV equations, which include the 
$A_\infty$ relations. Moreover, the formulas for topological correlators 
given in \cite{Kap2,Herbst:2004ax}, as well as the concrete study of 
the problem's deformation theory \cite{HW,dAADF}, have given valuable 
pieces of information about the above questions also in more geometrical 
settings. 

In the present paper, we study these problems for the simplest model 
that has a compact geometric interpretation, namely the cubic elliptic 
curve. The representation of its $B$-type branes in terms of matrix 
factorizations in the Landau-Ginzburg model has recently been discussed 
in \cite{HW}. It is based on the superpotential 
\beq
\eqlabel{cubicdef}
\wlg(x,\bpsi)\ =\ \frac13 {x_1}^3 + \frac13 {x_2}^3+\frac13 {x_3}^3 -
\bpsi\, x_1x_2x_3 \,,
\eeq
together with an obvious $\zet_3$ orbifold action. This model corresponds
to the point $\rho=\exp{2\pi\ii/3}$ in K\"ahler moduli space, and the
complex structure parameter varies as a certain function $\tau=\tau(\bpsi)$.
On the physics side, this model is exactly solvable at the CFT
level. On the mathematical side, the elliptic curve has been studied
extensively from the point of view of categorical mirror symmetry
in \cite{PoZa,Poli,Poli1,Poli2,Poli3,mirbook}, 
so that most of the questions we might want to ask should have known
answers. Our goal here is to learn how to derive some of these
results from the boundary Landau-Ginzburg realization, with the
expectation that the lessons we learn will be useful in more complicated
situations.

Specifically, we will focus on the computation of the effective
``Yukawa'' couplings associated with pairwise intersections of 
three branes. When expressed in flat coordinates, which we 
determine intrinsically in the $B$-model, these Yukawa couplings 
become the $A$-model generating functions for triangle-shaped
world-sheet instantons that span between the three $D$-branes.
From the point of view of categorical mirror symmetry, our results
amount to determining the associative Fukaya products $m_2$ from 
their Landau-Ginzburg $B$-model counterparts. The computation of 
the higher, non-associative products $m_k$ will be addressed 
elsewhere.

\section{$D$-branes, matrix factorizations and $Q$-cohomology}

As discussed in \cite{HW}, the $B$-type $D$-branes of this model can
be obtained from all possible matrix factorizations of \eqref{cubicdef}.
For $a=0$, those factorizations have been put in exact correspondence 
\cite{laza} with vector bundles on the elliptic curve $W=0\subset 
\IP^2$, which were classified by Atiyah. Simplest are the $\zet_3$-equivariant
$3\times 3$ factorizations involving the boundary BRST operators
\cite{HW}
\begin{equation}
\eqlabel{simpleQ}
Q_i = \begin{pmatrix} 0 & J_i \\ E_i & 0 \end{pmatrix}\ ,
\qquad i=1,2,3\ ,
\end{equation}
with
\begin{equation}
\eqlabel{simpleQQ}
\begin{split}
J_i &=
\begin{pmatrix}
{\alf1i} {x_1} & {\alf2i} {x_3} & {\alf3i} {x_2}\\
{\alf3i} {x_3} & {\alf1i} {x_2} & {\alf2i} {x_1} \\
{\alf2i} {x_2} & {\alf3i} {x_1} & {\alf1i} {x_3}
\end{pmatrix}
\\
E_i &=
\begin{pmatrix}
\frac{1}{{\alf1i}}{x_1}^2 - \frac{{\alf1i}}{{\alf2i}{\alf3i}} {x_2} {x_3} &
\frac{1}{{\alf3i}}{x_3}^2 - \frac{{\alf3i}}{{\alf1i}{\alf2i}} {x_1} {x_2} &
\frac{1}{{\alf2i}}{x_2}^2 - \frac{{\alf2i}}{{\alf1i}{\alf3i}} {x_1} {x_3} \\
\frac{1}{{\alf2i}}{x_3}^2 - \frac{{\alf2i}}{{\alf1i}{\alf3i}} {x_1} {x_2} &
\frac{1}{{\alf1i}}{x_2}^2 - \frac{{\alf1i}}{{\alf2i}{\alf3i}} {x_1} {x_3} &
\frac{1}{{\alf3i}}{x_1}^2 - \frac{{\alf3i}}{{\alf1i}{\alf2i}} {x_2} {x_3} \\
\frac{1}{{\alf3i}}{x_2}^2 - \frac{{\alf3i}}{{\alf1i}{\alf2i}} {x_1} {x_3} &
\frac{1}{{\alf2i}}{x_1}^2 - \frac{{\alf2i}}{{\alf1i}{\alf3i}} {x_2} {x_3} &
\frac{1}{{\alf1i}}{x_3}^2 - \frac{{\alf1i}}{{\alf2i}{\alf3i}} {x_1} {x_2}
\end{pmatrix}\ .
\end{split}
\end{equation}
The $\alf\ell i$ are parameters that are constrained by the
matrix factorization condition $Q_i^2(x,\alf\ell i)=W(x,\bpsi)\bfone$, 
which translates to \cite{HW}: 
\begin{equation}
\frac13(\alf1i)^3+\frac13(\alf2i)^3+\frac13(\alf3i)^3 -
\bpsi \,{\alf1i}{\alf2i}{\alf3i} = 0 \,.
\eqlabel{alpha}
\end{equation}
Thus, the moduli space spanned by the $\alf\ell i$ is isomorphic to the
Jacobian of the torus itself, and this is expected to hold for any
matrix factorization of \eqref{cubicdef}. As explained in \cite{HW},
the three particular matrix factorizations based on \eqref{simpleQ},
\eqref{simpleQQ} describe one-parameter deformations of the rational
$D$-branes, for any given value of the bulk modulus, $\bpsi(\tau)$.
These branes, which we shall denote by $\cL_1$, $\cL_2$, $\cL_3$,
are known \cite{laza} to correspond, in the geometric $B$-model category, to
bundles with ranks and degres given by $(r,c_1)=(2,1),(-1,1),(-1,-2)$,
respectively.\footnote{Their anti-branes are described by the equivalent 
factorizations obtained by swapping $E_i\leftrightarrow J_i$. Note also that we will
often denote branes and bundles by the same symbols $\cL_i$ in the
following.} In physics terms, these labels correspond to $D2$- and
$D0$-brane charges, respectively (the one-parameter deformations
correspond to the locations of the $D0$-branes on top of the
$D2$-branes, which by themselves wrap the cubic curve). 
Since $r+c_1=0\bmod 3$ for all three branes, the $\cL_i$ do not provide 
an integral basis of the complete $K$-charge lattice, which is a familiar 
feature in this context \cite{BDLR}. In the appendix, we exhibit a set of
$2\times 2$ matrix factorizations of the cubic that does correspond to
such an integral basis. 

\begin{figure}[htb]  
\begin{center}
\includegraphics[width=6cm]{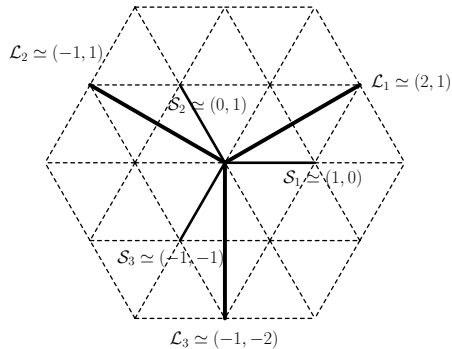}
\caption{Shown are the long and short diagonals on the covering
space of the torus; note that they correspond to roots and weights
of the $SU(3)$ lattice, resp. The long diagonals $\cL_i$ correspond,
via mirror symmetry, to the $3\times3$ matrix factorizarions
(\ref{simpleQQ}) we discuss in this paper, while the short diagonals
$\cS_i$ correspond to $2\times2$ factorizations.}
\label{branes}
\end{center}
\end{figure}

In the mirror description, in which (the roles of) $\rho$ and $\tau$
are exchanged, quasihomogeneous matrix factorizations correspond
to branes wrapped along special Lagrangian submanifolds of the torus
$\complex/(\zet+\rho\zet)$, with wrapping numbers $(n_1,n_2)=(r,c_1)$. 
In particular, the $A$-model mirrors of the three branes $\cL_i$ 
described by \eqref{simpleQQ} can be pictured as the three long 
diagonals of the $SU(3)$ torus, see Fig.\ \ref{branes}. In $A$-model 
language, the boundary moduli correspond to position
and flat gauge fields on the lines, and we will describe further
below the mirror map between them and the $B$-model moduli
$\bpsi$, $\alf \ell i$.

\begin{figure}[htb]  
\begin{center}
\includegraphics[width=6cm]{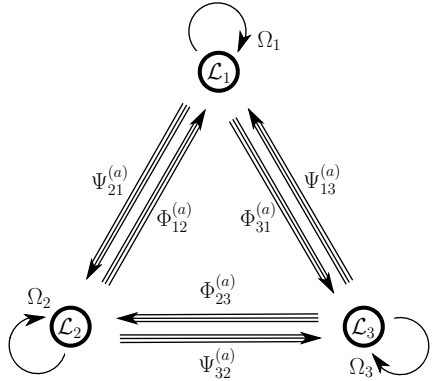}
\caption{Quiver representation of the open string spectrum between
the three $D$-branes $\cL_i$ under consideration. One of our objectives is to
find suitable Landau-Ginzburg representatives of all the pictured
quantities that continuously depend on the bulk/boundary moduli.
}
\label{quiver}
\end{center}
\end{figure}

We now turn to discussing the boundary changing operators, that is,
cohomology representatives of the open string spectrum
between pairs of the $\cL_i$. We have summarized the open string
spectrum in the quiver diagram of Fig.\ \ref{quiver}. As indicated,
in the boundary changing sector between $\cL_j$ and $\cL_i$ (with 
$i=j+1\bmod 3$) there are three bosonic and three fermionic elements, 
$\Phi_{ji}\alb a$ resp.~$\Psi_{ij}\alb a$ ($a=1,2,3$). These correspond 
to the three intersection points each pair of branes has, when translated 
to a fundamental domain. Moreover, $\Omega_i$ denote boundary preserving
operators of top degree (R-charge) $1$, which generate the marginal
deformations of the branes. The $\Omega_i$ will be discussed at 
length in the next section.

In order to construct LG representatives, it is useful to first determine the
degrees (charges) of the open string operators. Note that \eqref{simpleQQ} 
is quasihomogeneous with R-charge assignement
\begin{equation}
\ee^{\ii\lambda R} = 
\begin{pmatrix} \ee^{\ii\lambda/6} \,\bfone_3 & 0 \\ 0 & \ee^{-\ii\lambda/6}\,
\bfone_3
\end{pmatrix}
\end{equation}
and equivariant with respect to the following orbifold action
\begin{equation}
\ee^{2 \pi j \ii/3} 
\begin{pmatrix} \bfone_3 & 0 \\ 0 & -\ee^{-\ii\pi/3}\,
\bfone_3
\end{pmatrix}  \qquad (j=1,2,3)
\end{equation}
on the Chan-Paton spaces. Therefore, in order to survive the 
orbifolding, $\Phi_{ij}\alb a$ and $\Psi_{ij}\alb a$ must have 
R-charge $q_\Phi=2/3$ and $q_\Psi=1/3$, respectively. [Note,
however, that this is subject to change once we move the K\"ahler 
modulus away from $\rho=\exp{2\pi\ii/3}$. The important invariant 
statement is $q_\Phi+q_\Psi=1$ by charge conjugation
(Serre duality), and $0<q_\Phi<1$ so that $\Phi$ and $\Psi$ are
always tachyonic; there are no lines of marginal stability on the
torus.]

We start with finding representative of the fermionic operators 
$\Psi_{ij}\alb a$ mapping from $\cL_j$ to $\cL_i$. We will explicitly 
take $i=2$ and $j=1$, but everything works analogously for $i=j+1\bmod 3$.
Writing
\begin{equation}
\Psi_{21} = \begin{pmatrix} 0 & F_{21} \\ G_{21} & 0 \end{pmatrix}\,,
\end{equation}
$Q$-closedness requires that $F$ and $G$ satisfy
\begin{equation}
\begin{split}
J_2 G_{21} + F_{21} E_1 &= 0 \\
E_2 F_{21} + G_{21} J_1 & = 0
\end{split}
\eqlabel{coho}
\end{equation}
The above degree considerations in the orbifold dictate that $F$ be 
constant (\ie, independent of ${x_\ell}$) and $G$ be linear in ${x_\ell}$.
One may also note that the image of the $Q_i$'s at this degree is zero
(there are no bosonic operators in degree $-2/3$, as this would
require negative powers of $x_\ell$), so that all solutions to
\eqref{coho} will be cohomologically non-trivial.

All-in-all, one indeed finds three linearly independent solutions of
\eqref{coho}, which are precisely the $\Psi_{21}\alb a$ we are looking
for. The first one reads
\begin{equation}
\eqlabel{FG}
F^{(1)}_{21} = \begin{pmatrix}
{\zta1{}} & 0 & 0 \\ 0 & 0 & {\zta2{}} \\ 0 & {\zta3{}} & 0 \end{pmatrix} 
\qquad
G^{(1)}_{21} = -\begin{pmatrix}
\frac{{\zta1{}}}{{\alf11}{\alf12}} {x_1} & 
\frac{{\zta3{}}}{{\alf11}{\alf22}} {x_2} &
\frac{{\zta2{}}}{{\alf11}{\alf32}} {x_3} \\
\frac{{\zta2{}}}{{\alf12}{\alf31}} {x_2} &
\frac{{\zta1{}}}{{\alf22}{\alf31}} {x_3} &
\frac{{\zta3{}}}{{\alf31}{\alf32}} {x_1} \\
\frac{{\zta3{}}}{{\alf12}{\alf21}} {x_3} & 
\frac{{\zta2{}}}{{\alf21}{\alf22}} {x_1} &
\frac{{\zta1{}}}{{\alf21}{\alf32}} {x_2}
\end{pmatrix}\ .
\end{equation}
Inserting this ansatz into \eqref{coho} results in $18$ equations, out of which only
two are independent if \eqref{alpha} is used, \eg,
\begin{equation}
\begin{split}
\frac{{\zta1{}}{\alf12}}{{\alf22}{\alf32}} + \frac{{\zta2{}}{\alf21}}{{\alf11}{\alf32}}
+\frac{{\zta3{}}{\alf31}}{{\alf11}{\alf22}} &= 0 \\
\frac{{\zta1{}}{\alf21}}{{\alf11}{\alf12}} + \frac{{\zta2{}}{\alf31}}{{\alf11}{\alf32}}
+\frac{{\zta3{}}{\alf22}}{{\alf12}{\alf32}} &=0\ .
\end{split}
\end{equation}
\begin{equation}
\eqlabel{zetadef}
\begin{split}
{\zta1{}} &= (\alf22)^2 {\alf11} {\alf21} - {\alf12}{\alf32}(\alf31)^2 \\
{\zta2{}} &= (\alf32)^2 {\alf21}{\alf31} - {\alf12}{\alf22}(\alf11)^2 \\
{\zta3{}} &= (\alf12)^2 {\alf11}{\alf31} - {\alf22}{\alf32}(\alf21)^2\ .
\end{split}
\end{equation}
One may note that the ${\zta \ell{}}$ also satisfy the cubic equation
\begin{equation}
\eqlabel{zetacubic}
\frac13 {\zta1{}}^3+\frac13{\zta2{}}^3+\frac13{\zta3{}}^3-\bpsi\,{\zta1{}}{\zta2{}}{\zta3{}}=0
\end{equation}
which identifies $({\zta1{}}$, ${\zta2{}}$, ${\zta3{}})$ as a point on the 
(Jacobian of the) torus; this also follows upon inserting the ansatz into 
\eqref{coho} and taking determinants.

The second and third solutions take the form:
\begin{equation}
\eqlabel{FG23}
\begin{split}
F^{(2)}_{21}&=\begin{pmatrix} 0 & 0 & {\zta3{}} \\ 0 & {\zta1{}} & 0\\ {\zta2{}} & 0 & 0
\end{pmatrix} 
\qquad
G^{(2)}_{21}= - \begin{pmatrix}
\frac{{\zta1{}}}{{\alf21}{\alf32}} {x_3} &
\frac{{\zta3{}}}{{\alf12}{\alf21}} {x_1} &
\frac{{\zta2{}}}{{\alf21}{\alf22}} {x_2} \\
\frac{{\zta2{}}}{{\alf11}{\alf32}} {x_1} &
\frac{{\zta1{}}}{{\alf11}{\alf12}} {x_2} &
\frac{{\zta3{}}}{{\alf11}{\alf22}} {x_3} \\
\frac{{\zta3{}}}{{\alf31}{\alf32}} {x_2} &
\frac{{\zta2{}}}{{\alf12}{\alf31}} {x_3} &
\frac{{\zta1{}}}{{\alf22}{\alf31}} {x_1}
\end{pmatrix}\\
F^{(3)}_{21}&=\begin{pmatrix} 0 & {\zta2{}} & 0 \\ {\zta3{}} & 0 & 0\\0 & 0 & {\zta1{}}
\end{pmatrix} 
\qquad
G^{(3)}_{21}= - \begin{pmatrix} 
\frac{{\zta1{}}}{{\alf22}{\alf31}}{x_2} &
\frac{{\zta3{}}}{{\alf31}{\alf32}} {x_3} &
\frac{{\zta2{}}}{{\alf12}{\alf31}} {x_1} \\
\frac{{\zta2{}}}{{\alf21}{\alf22}} {x_3} &
\frac{{\zta1{}}}{{\alf21}{\alf32}} {x_1} &
\frac{{\zta3{}}}{{\alf12}{\alf21}} {x_2} \\
\frac{{\zta3{}}}{{\alf11}{\alf22}} {x_1} &
\frac{{\zta2{}}}{{\alf11}{\alf32}} {x_2} &
\frac{{\zta1{}}}{{\alf11}{\alf12}} {x_3}
\end{pmatrix}\ ,
\end{split}
\end{equation}
respectively, with the same values of ${\zta \ell{}}$ as above. 
These three solutions 
correspond precisely to the threefold arrows in the quiver diagram Fig.\ 
\ref{quiver} that can be associated with the ambient space geometry.

The arrows pointing in the opposite direction also come triply
degenerate, and correspond to bosonic boundary ring elements
$\Phi_{ji}^{(a)}$, $a=1,2,3$ (with $i=j+1\bmod 3$). 
Their matrix representations
are block diagonal with both blocks linear in the ${x_\ell}$, and
depend on a choice of gauge because the image of $Q_i$'s at degree
$2/3$ is non-trivial.  Of course, as for the fermions, we have in
mind a basis with a definite ``triality'', \ie, we require that
$\Psi_{21}^{(a)}$ and $\Phi_{12}^{(a)}$ are Serre dual to each
other. These considerations lead to the ansatz
\begin{equation}
\Phi_{12}^{(1)} = \begin{pmatrix} H^{(1)} & 0 \\ 0 & K^{(1)} \end{pmatrix}
\end{equation}
with
\begin{equation}
H^{(1)} =
\begin{pmatrix}
h_{11} x & h_{12} y & h_{13} z \\
h_{21} z & h_{22} x & h_{23} y \\
h_{31} y & h_{32} z & h_{33} x
\end{pmatrix} 
\qquad
K^{(1)} =
\begin{pmatrix}
k_{11} x & k_{12} z & k_{13} y \\
k_{21} y & k_{22} x & k_{23} z \\
k_{31} z & k_{23} y & k_{33} x
\end{pmatrix}
\end{equation}
Solving 
\begin{equation}
J_1 K^{(1)} -  H^{(1)} J_2 = 0 \,;
\qquad E_1 H^{(1)} -  K^{(1)} E_2 = 0
\eqlabel{boscoho}
\end{equation}
modulo
\begin{equation}
\delta H^{(1)} = J_1 L \,; \qquad
\delta K^{(1)} = L J_2
\eqlabel{exact}
\end{equation}
where $L$ is an arbitrary scalar matrix, yields the following solution,
in the simplest gauge we could find: 
\def\aalf(#1,#2){\alf{#1}{#2}}
\def\xxi(#1){x_{#1}}
\def\zetb{\bar\zeta}
\beql{bosonic}
H^{(1)}\ =\ 
\left(\begin{matrix}
 0 & \frac{\aalf(3,1)\,\aalf(3,2)\,\xxi(2)\,\zeta_3}{\aalf(1,2)} & \frac{\aalf(2,1)\,
     \aalf(2,2)\,\xxi(3)\,\zetb_3}{\aalf(1,2)} \cr 
     \frac{\aalf(2,1)\,\aalf(3,1)\,\aalf(3,2)\,
     \xxi(3)\,\zetb_3}{\aalf(1,1)\,\aalf(1,2)} & 0 & \aalf(2,1)\,\xxi(2)\,
   \zeta_3 \cr 
   \frac{\aalf(2,1)\,\aalf(2,2)\,\aalf(3,1)\,\xxi(2)\,\zeta_3}
   {\aalf(1,1)\,\aalf(1,2)} & \aalf(3,1)\,\xxi(3)\,\zetb_3 & 0 \cr  
   \end{matrix}\right)\ ,
   \\
    K^{(1)}\ =\   
\left(\begin{matrix}
    0 & \frac{\aalf(2,1)\,\aalf(2,2)\,\aalf(3,2)\,\xxi(3)\,\zetb_3}
   {\aalf(1,1)\,\aalf(1,2)} & \frac{\aalf(2,2)\,\aalf(3,1)\,\aalf(3,2)\,\xxi(2)\,\zeta_3}
   {\aalf(1,1)\,\aalf(1,2)} \cr \frac{\aalf(2,1)\,\aalf(2,2)\,\xxi(2)\,\zeta_3}
   {\aalf(1,1)} & 0 & \aalf(2,2)\,\xxi(3)\,\zetb_3 \cr \frac{\aalf(3,1)\,\aalf(3,2)\,\xxi(3)\,
     \zetb_3}{\aalf(1,1)} & \aalf(3,2)\,\xxi(2)\,\zeta_3 & 0 \cr
   \end{matrix}\right)\ ,  \nonumber       
\eeql
where $\zetb_3$ is like $\zeta_3$ in (\ref{zetadef}), 
except that $\alf i1$ and $\alf i2$ are exchanged.
The other bosonic operators $\Phi_{ij}^{(a)}$ with $a=2,3$ can be 
similarly dealt with, and we refrain from presenting them here.

\goodbreak

\section{Flat coordinates of brane-bulk moduli space}
\label{flat}

A crucial piece of mirror symmetry is the map between the algebraic
coordinates the $B$-model and the flat ``geometric'' coordinates,
which are natural in the $A$-model. Due to the simplicity of the
torus, we know the answer beforehand: the flat coordinates are given by the
complex structure parameter $\tau$ of the curve (which under mirror
symmetry becomes identified with the K\"ahler parameter $\tilde\rho$ of
the dual torus), and the brane locations $u_i$, living on the
jacobian which is isomorphic to the torus itself (in the $A$-model
picture, the $u_i$ are complex variables that combine shift and
Wilson line moduli).

In fact it is known since a long time \cite{Klein} what the functions
$\bpsi$ and $\al_\ell$ are in terms of $\tau$ and $u$.  Specifically,
the algebraic modulus $\bpsi$ is related to the flat complex structure
modulus $\tau$ as a modular function for $\Gamma[3]$, defined via the
following relationship to the modular invariant~$J(\tau)$:
\beq
\eqlabel{Jpsimap}
\left({\frac{J(\tau)}{1728}}\right)^{1/3}\ 
=\ -\frac1{4}\frac{\bpsi(\bpsi^3+8)}{1-\bpsi^3}\ .
\eeq
Moreover, the $\al_\ell$ are given by certain Weierstrass
$\sigma$-functions, which coincide (up to a common prefactor) with
Jacobi $\tht$-functions evaluated at third-points. The underlying
mathematical reason is that the $\tht$-functions 
($q\equiv e^{2\pi i\tau}$):
\beql{THETAdef}
\tht\Big[{\topa{c_1}{c_2}}\Big|\,n\, u,n\tau\Big]\ =\
\sum_m q^{n(m+c_1)^2/2} e^{2\pi i(n\, u+c_2)(m+c_1)}\ ,
\eeql
for $c_2=0$, $c_1=k/n$ ($k=0,...,n-1$), form a basis of global sections
of degree $n$ line bundles $L(n,u)\cong L^{\otimes (n-1)}L(u)$, and
provide a projective embedding of the elliptic curve. From the cubic
representation of the curve it follows that we need to take $n=3$.
Moreover, what we are after are sections of the sheaf $\cO(u_i)$
of holomorphic functions whose zeros are at the values of the
boundary moduli $u_i$. Since $\cO(u)\simeq L(u-u_0)$
where $u_0=\frac{1+\tau}2\,{\rm mod}\,\ZZ\times\tau\ZZ$,
we shift the characteristics of the $\tht$-functions by $-1/2$.
 
Apart from normalization, there is a further ambiguity in identifying the 
$\al_\ell$ with these $\tht$-functions, and this reflects the action of 
the monodromy group which is given by the tetrahedral group, 
$\cT=\Gamma/\Gamma[3]$. Like $\bpsi(\tau)$, the $\al_\ell$ transform 
under the action of $\cT$ (as has been discussed in \cite{LLW}, the LG 
fields ${x_\ell}$ transform as well, and presumably also the Chan-Paton
matrices). We fix the ambiguity such that $\al_1\rightarrow 0$ if we 
approach the Gepner point $\bpsi=0$, which are the conventions used in 
\cite{HW}.  We thus identify, up to a common normalization:
\beq
\eqlabel{alphaTHETA}
\alf\ell i\ \equiv\ \ \alf\ell i(\tau,u_i)\ =\
\epsilon^\ell\, \tht\Big[{\topa{(1-\ell)/3-1/2}{-1/2} }\,\Big|\,3u_i,3
\tau\Big]\ ,\qquad \ell=1,2,3,
\eeq
where $\epsilon=\ii e^{2\pi\ii/3}$.  As we have mentioned,
the index labels lattice conjugacy classes, and thus $(\ell-1)$ can
be viewed as a $\zet_3$-valued ``charge'' that is preserved under
multiplication. Using \eqref{Jpsimap}, it is easy to check that the
$\alf\ell i(u_i,\tau)$ indeed satisfy the cubic relation 
(\ref{alpha}).\footnote{We have to choose the proper branch of 
$\bpsi(J(\tau))$ that matches our choice of $\al_\ell$'s, and we find 
that the correct choice is given by the branch that goes like $q^{-1/3}$.}

We now like to identify a flat basis of the bulk/boundary cohomology
representatives corresponding to $\tau$ and $u$ directly from LG
considerations.  By definition, marginal deformations come from
derivatives of the LG potentials. In the bulk sector we will
take as usual $\phi(x,\tau)= -\del_\tau W(x,\tau)$, while on the
boundary we are lead to consider:
\beq
\eqlabel{flatOmega}
\Omega(x,\tau,u)\ =\ \frac\del{\del u}\,Q(x,\alpha_\ell(\tau,u))\ .
\eeq
This is BRST invariant due to
$\frac12\{\Omega,Q\}=\del_u\{Q,Q\}=\del_uW=0$.  The ansatz
(\ref{flatOmega}) can be justified by either one of the following
two interrelated chains of arguments. We just outline the first one
(which is based on the variation of Hodge structures), because the
second one (based on constancy of the topological metric) is much
easier to spell out in the present situation.

First, one may derive differential equations for
an appropriate generalized period integral involving $\Omega$, the
solutions of which will determine the flat coordinates in a systematric
way. A natural integral over fermionic variables is given by
$\str[Q\,\cdot\,]W^{-1}$, and we thus may consider variations
of\footnote{Equivalently, we also could consider $\tilde\Pi=\int
d^3x\,\str[\Omega e^{-Q}]e^{-W}$.}
\beql{bperiod} 
\Pi_0^\al\
=\ \int_{\gamma_\al}\lambda\ ,
\qquad
\lambda\ =\ \int_{\gamma_W}
\frac{\str[Q\,\Omega]}{W(x,\bpsi)^2}\,\omega\  , 
\eeql 
where $\omega=\sum_{\ell=1}^{3}(-1)^\ell
x^\ell\,dx^1\!\wedge\!\dots\wedge\!{\widehat{dx^\ell}}
\!\wedge\!\dots\!\wedge\!dx^{3}$ is a volume element, and $\gamma_W$
is a small loop around the locus $W=0$ in $\IP^2$.  Similar as
explained in \cite{LSW}, a flat basis is characterized by the
vanishing of double derivatives of $\Pi_0^\al$.
   This can be achieved
by requiring that the supertrace maps $\lambda$ to the holomorphic
1-form $\eta=\int \omega/W$ on the curve, which maps the problem
to an already solved one. Indeed, $\Omega$ in (\ref{flatOmega}) has
the key property that
\beq
\eqlabel{bflatness}
\str[Q\,\Omega](x)\big|_{\del_\ell W(x)=0}\ =\ 0\ ,
\eeq
so that all contributions to the period integral come from ``contact''
terms that are proportional to derivatives of $W(x)$.  Upon integrating
by parts and choosing an appropriate normalization factor, the
$\Pi_0^\al$ for $\al=0,1$ can then be made to coincide with the
ordinary periods associated with the torus, if we choose for
$\gamma_{0,1}$, the usual symplectic homology basis of
1-cycles on the elliptic curve. 

Moreover, we also introduce a 1-chain $\gamma_2$ in the relative
homology, one boundary of which sits at a point $p$ of the elliptic
curve ($p$ can be interpreted as the location of a $D0$-brane on
the $T^2$; this is analogous to the considerations of ref.~\cite{LMW},
where 3-chains on Calabi-Yau threefolds where considered whose
boundaries are the locations of $D2$-branes).  The line integral
over the chain $\gamma_2$ will give an extra, functionally independent
semi-period, associated with the open string modulus.

Following the arguments of \cite{LMW}, we know that the $\Pi^\al_0$
must satisfy a system of differential equations that will determine
the flat cordinates.  However, these turn out to be very complicated to
write down and solve in terms of a general matrix ansatz for $\Omega$
and the LG variables $\al_\ell$ and $\bpsi$. On the other hand,
since we know the flat coordinates $\tau,u$ anyway, we can express
$\Omega$ as given in (\ref{flatOmega}) in terms of them and compute
the differential equations and their solutions directly in the flat
coordinates. Concretely, after some lengthy calculations, this
yields the following simple linear system:
\beql{flatPF}
\left[{\frac{\del}{\del\tau}}\,-\,
\left(\begin{matrix}
 0 & 1 & 0\\
  0 & 0 & 0\\
 0 & 0 & 0\\ 
\end{matrix}\right)\right]\cdot
\Pi(\tau,u)
\ =\ 0\ ,\qquad
\left[{\frac{\del}{\del u}}\,-\,
\left(\begin{matrix}
 0 & 0 & 1\\
 0 & 0 & 0\\
 0 & 0 & 0\\ 
\end{matrix}\right)\right]\cdot
\Pi(\tau,u)
\ =\ 0\ ,
\eeql
which is trivially satisfied by the relative period matrix:
\beql{periodm}
\Pi_\be^\al(\tau,u)\ =\ 
\left(\begin{matrix}
q(\tau)\int_{\gamma_\al}\eta \\
\frac\del{\del\tau}
q(\tau)\int_{\gamma_\al}\eta \\
\frac\del{\del u}
q(\tau)\int_{\gamma_\al}\eta \\
\end{matrix}\right)
\ =\ 
\left(\begin{matrix}
1 & \tau & u\\
0 & 1 & 0\\
0 & 0 & 1\\
\end{matrix}\right)\ .
\eeql
Here, 
\beql{flattenq}
q(\tau)\ =\ \left(\frac {1-\bpsi^3}{3\bpsi'(\tau)}\right)^{\frac12}\ ,
\eeql
is a ``flattening'' normalization factor \cite{LSW} that is needed
in order to get rid of all the connection terms in the matrix
differential equations. This factor can be understood as a particular
change of normalization\footnote{On a Calabi-Yau
threefold, one would refer to this as a canonical choice of K\"ahler gauge.
It amounts to dividing out the periods by the unique period that
behaves as a power series at large $\tau$.} of the bulk potential: 
$W\to  q(\tau)^{-1}W$ (or equivalently, of the holomorphic one-form).  

A much more direct way to show that $u$ is a flat coordinate and
$\Omega$ as given in (\ref{flatOmega}) is a good flat cohomology
representative, is given  by computing the topological metric in the
boundary sector, and verifying it to be constant. For this, it is
important to note that the factorization condition $Q^2=W$ constrains
the relative normalization of $Q$ and $W$.  In particular, the
flattening factor for $Q$ must be $q(\tau)^{-1/2}$ and this cannot
depend on the boundary parameters. Therefore, flatness of $u$ should
be equivalent to constancy of the boundary topological metric, \ie,
of the disk correlator $\langle\Omega\rangle_{\rm disk}$ when using the correct
normalization of $W$. Indeed, by plugging \eqref{flatOmega} into
the generalized residue formula for topological correlators of
\cite{Kap2,Herbst:2004ax}, we find by direct computation
\begin{equation}
\eqlabel{disk}
\begin{split}
\langle\,\Omega\,\rangle_{\rm disk, normalized}
&\equiv\
q(\tau)\int\frac{\str[\frac1{3!}(dQ)^{\wedge3}\del_{u} Q]}
{\del_{1}W\del_{2}W\del_{3}W(x)} \\
&=\ \int\frac{f(\tau,u)\,H(x)}{\del_{1}W\del_{2}W\del_{3}W(x)}
\ =\ f(\tau,u)\ ,
\end{split}
\end{equation}
with
\begin{equation}
\eqlabel{result}
f(\tau,u)\ =\ 
q(\tau)\frac{\frac1{2\pi i}\del_{u}\alf12(\tau,u)}
{\alpha_2(\tau,u)^2-\bpsi(\tau)\,\alf12(\tau,u)}\ .
\end{equation}
In \eqref{disk}, $H(x)={\rm det}\del_i\del_jW(x,\bpsi)$ is the hessian of the
superpotential whose residue integral equals unity.  

Imposing $\langle\Omega\rangle_{\rm disk, normalized}=1$ is equivalent to the 
statement that $u$ is a coordinate of the jacobian, which is what is expressed 
in \eqref{alphaTHETA}. Indeed, the holomorphic one-form on the cubic curve 
described by \eqref{alpha} looks in the local patch $\al_3=1$ as
\begin{equation}
\eta = q(\tau)\frac{ d\al_1}{\partial_{\al_2} W(\al_1,\al_2,1)}
= q(\tau)\frac{d \al_1}{{{\al_2}}^2 - \bpsi\al_1}\ .
\end{equation}
Therefore $f(\tau,u)=1$ is solved by
\begin{equation}
u\  = \ \int_{p_0}^{p(u)} \eta\ ,
\end{equation}
where $p(u)=\alpha_1(u)$ and $p_0$ is some reference point which
we take to be $\infty$. This identifies $u$, defined via
$\langle\Omega\rangle\equiv\langle\del_u Q\rangle=1$, as a
flat coordinate on the jacobian, as expected.

\section{Boundary changing correlators and disk instantons}

We now turn to determining correlation functions. We just have seen
that in the sector of a single $D$-brane, the disk correlator
$\langle\,\Omega\,\rangle$ is non-zero. However, this does not imply
that there is a non-zero effective superpotential.  This topological
correlator corresponds to a boundary 3-point function
$\langle\,1\,1\,\Omega\,\rangle$, but the insertions of the boundary
identity operator do not correspond to taking derivatives of an
effective potential with respect to moduli.\footnote{Rather, these
operators correspond to formal fermionic deformation parameters,
which cancel out in the effective potential \cite{HLL}). One may
also view $\langle\,\Omega\,\rangle$ as a 2-point function, but
again the identity operator does not correspond to a modulus in the
effective action.}
 That there is no effective superpotential generated in the boundary
 preserving sector of a single $D$-brane reflects, of course, that
 the deformations parametrized by $\tau$ and $u$ are not obstructed.

In order to obtain a non-trivial superpotential, we thus need to
resort to correlators of boundary changing operators, and we will
specifically consider 3-point functions of the form:
\beql{threepoint}
\langle\,\Psi_{13}^{(a)}\Psi_{32}^{(b)}\Psi_{21}^{(c)}\,\rangle\ =\
\langle\,C_{abc}\Omega_1\,\rangle\ =\ C_{abc}(\tau,u_1,u_2,u_3)\ ,
\eeql
which correspond to going around once in the quiver diagram of Fig.~\ref{quiver}.
Here $\Psi_{ij}^{(a)}$ denotes the fermionic ring elements of Section~2,
which correspond to open strings streching between the
$D$-branes $\cL_j$ and $\cL_i$. Their proper
normalization still needs to be determined.

Let us parametrize the normalization of the boundary ring elements
by a priori unknown functions $g=g(\tau,u)$, and write the
full BRST operator in the following way:
\beql{fullQ}
\cQ\ =\ \left(\begin{matrix}
Q_1(\tau,u_1) & \sum t_{12}^{(a)}g_{12}^{(a)}\Phi_{12}^{(a)} & \sum t_{13}^{(a)}g_{13}^{(a)}\Psi_{13}^{(a)}\\
\sum t_{21}^{(a)}g_{21}^{(a)}\Psi_{21}^{(a)} & Q_2(\tau,u_2) & \sum t_{23}^{(a)}g_{23}^{(a)}\Phi_{23}^{(a)} & \\
\sum t_{31}^{(a)}g_{31}^{(a)}\Phi_{31}^{(a)} & \sum t_{32}^{(a)}g_{32}^{(a)}\Psi_{32}^{(a)} & Q_3(\tau,u_3) \\
\end{matrix}\right)\ ,
\eeql
where $t^{(a)}_{ij}$, $a=1,2,3$ are the triplets of tachyon fields
between the branes $\cL_j$ and $\cL_i$ that are defined by
 $\frac{\del}{\del t^{(a)}_{ij}}\cQ=
g_{ij}^{(a)}\Psi_{ij}^{(a)}$. When they take generic
values, the matrix factorization $\cQ\cdot\cQ=W\bfone$ is spoiled, and
this reflects that deformations along these directions are
generically obstructed. In other words, there will be a non-vanishing effective
superpotential $\weff$ of the form\footnote{Note that the ordering
of the $t$'s is important here, and one may prefer to treat them
as non-commuting quantities.}
\beql{weffdef}
\weff(t,\tau,u_i)\ =\ \sum_{j>k>i\,{\rm mod}\ 3}  C_{abc}(\tau,u_i)
t^{(a)}_{ij}t^{(b)}_{jk}t^{(c)}_{ki}\ .
 + \cO(t^4)\
\eeql
As indicated, there are higher order corrections
in the tachyons, and specifically another term allowed by charge conservation is 
$t^{(a)}_{12}t^{(b)}_{23}t^{(c)}_{31}t^{(d)}_{12}t^{(e)}_{21}$.
Presumably it can be determined by making
use of the generalized consistency conditions (which include the
$A_\infty$ relations) derived in ref.\ \cite{HLL}. However, our
purpose in this paper is to just determine the 3-point functions
$C_{abc}(\tau,u_i)$ in terms of the unobstructed deformation
parameters.

For obtaining the proper normalization, one might at first want to require
the constancy of the topological 2-point functions (which reflect
Serre duality). However, the
trace structure of disk correlators implies that it is only the
product of both fermionic and bosonic normalization functions that
is constrained in this way,
\beql{topmetric}
\qquad
g_{ij}^{(a)}(\tau,u)g_{ji}^{(b)}(\tau,u)\,\langle\,\Psi_{ij}^{(a)}
\Phi_{ji}^{(b)}\,\rangle
\  \ \mathop{=}^!\ \ \delta^{ab}\ ,
\eeql
and this does not help us determining the absolute normalization
of the fermionic 3-point functions (\ref{threepoint}).

To proceed, let us first simplify the expressions for the
$\Psi_{ij}^{(a)}$ given in Section~2. Recall that the functions
$\zeta_\ell$ also satisfy the cubic equation, cf., (\ref{zetacubic}),
and thus also should be given by $\tht$-functions.  It turns
out, as a consequence of the quartic addition formulas \cite{tata}
that the $\tht$-functions obey, that
\beql{zetaident}
 \zeta_\ell(u_i,u_j)\ =\ c_{ij}\,\al_\ell(-u_{i}-u_j)\ ,
\eeql
where $c_{ij}=\eta^2\al_3(u_j-u_i)$ is independent of $\ell$.  
Thus, by a change of overall
normalization, we will take as a new ring basis the matrices
$\Psi_{ij}^{(a)}$ as described before, but now with the substitutions
 $\zeta_\ell\rightarrow\al_\ell(-u_i-u_j)$.  We will see later in
 Section~5 that this way of writing the $\Psi$'s is more natural
 from the mathematical point of view. Moreover, as we will see
 momentarily, the normalization of the three-point correlators will
 be already very close to the correct result.
 
 The result depends on which of the three kinds of the open string
 intermediate states are considered. One can associate a 
 $\ZZ_3$-valued charge associated with the label $(a)$, and there is a
 selection rule which requires that the total $\ZZ_3$ charge of any
 correlator must vanish. All-in-all there are only three
 independent kinds of non-vanishing correlators. Specifically,
 we find after somewhat cumbersome calculations that the $\tht$-functions
very nicely conspire such that the complicated expressions for
 the correlators collapse to the following
simple ones (see the next section for a rationale):
\beql{cijkresult}
C_{111}(\tau,u_i)\ &\sim&\  \frac{q(\tau)}{\eta(\tau)} \,\al_1(\tau,u_1+u_2+u_3)\nonumber\\
C_{123}(\tau,u_i)\ &\sim&\  \frac{q(\tau)}{\eta(\tau)} \,\al_2(\tau,u_1+u_2+u_3)\\
C_{132}(\tau,u_i)\ &\sim&\  \frac{q(\tau)}{\eta(\tau)} \,\al_3(\tau,u_1+u_2+u_3)\ .\nonumber
\eeql
In order to fix the overall normalization, we now make use of the 
following operator product:
\beql{OmegaSquare}
\Omega_i(x,\tau,u_i)\cdot \Omega_i(x,\tau,u_i)\ 
=\ 12 \pi i\,{\bf 1}\,\phi(x,\tau) 
\ \ \mathrm{mod.}\ \ \del_\ell W(x,\tau)\ ,
\eeql
which can be verified by direct computation. 
Note that despite the marginal bulk
operator $\phi(x,\tau)$ does not belong to the
boundary cohomology, integrated insertions of it in correlators can
still contribute at the boundary via contact terms. Because the operator 
identity (\ref{OmegaSquare}) involves the ring elements in a flat basis, 
it imposes the following simple derivative, ``Ward-identity'' on correlators:
\beql{wardid}
\left(\frac{\del^2}{\del {u_i}^2}-12\pi i
 \frac\del{\del\tau}\right)\,C_{abc}(\tau,u_i)\ =\ 0.
\eeql
This is nothing but the one-dimensional heat equation which is known
to be satisfied by $\tht$-functions \cite{tata}; in fact, it is
satisfied precisely by the $\tht$-functions that define the sections
$\al_\ell$ in (\ref{alphaTHETA}).  In other words, the correct
normalization of the correlators is given (up to a constant) just
by the expressions (\ref{cijkresult}) with the common prefactors
dropped. 

\begin{figure}[htb]  
\begin{center}
\includegraphics[width=9cm]{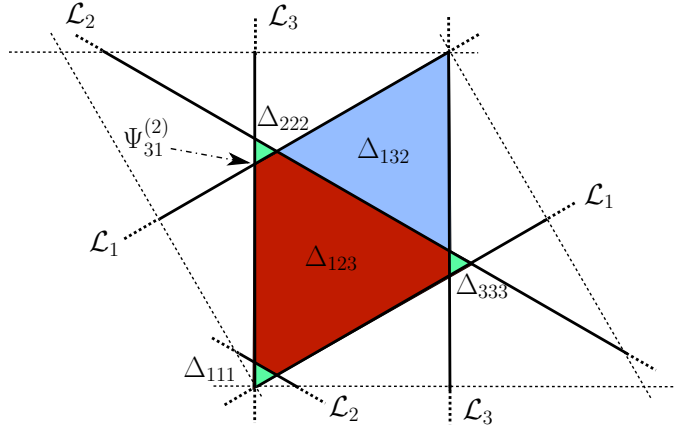}
\caption{Shown is the fundamental region of the cubic torus at $\rho=e^{2\pi i/3}$, 
with the three special lagrangian
$D$-branes $\cL_i$ on top.  The triangular world-sheets 
$\Delta_{abc}$ shown give the leading instanton corrections to the Yukawa
couplings $C_{abc}$.  Note that we have slightly shifted ${\mathcal
L}_2$ by setting $u_2\not=0$, so that each of the
three triple intersections gets resolved into three pairwise
intersections, and the $\Delta_{aaa}$ get a non-vanishing area.
The boundary changing open string operators $\Psi_{ij}^{(a)}$
are localized at the corresponding intersection points of the branes
$\cL_j$ and $\cL_i$ (an example of which is indicated).
}
\label{triangles}
\end{center}
\end{figure}

Now recall that our parametrization of the jacobian in (\ref{alphaTHETA})  
was such that we had switched on certain Wilson lines and position shifts.
Undoing these translations (the choice of origin on the jacobian
is of course immaterial), we finally obtain for the 3-point functions:
\beql{cijk}
C_{111}(\tau,\xi)\ &=&\  
e^{6\pi i \xi_1\xi_2}q^{3{\xi_2}^2/2}\,
\sum_m q^{3m^2/2} e^{6\pi i m\xi } \nonumber\\
C_{123}(\tau,\xi)\ &=&\  
e^{6\pi i \xi_1\xi_2}q^{3{\xi_2}^2/2}\,
\sum_m q^{3(m+1/3)^2/2} e^{6\pi i  (m+1/3)\xi } \\
C_{132}(\tau,\xi)\ &=&\  
e^{6\pi i \xi_1\xi_2}q^{3{\xi_2}^2/2}\,
\sum_m q^{3(m-1/3)^2/2} e^{6\pi i (m-1/3)\xi  } \ ,\nonumber
\eeql
where $\xi\equiv\xi_1+\tau\xi_2= u_1+u_2+u_3$.
In $A$-model language where $\tau\rightarrow\rho$,
the interpretation \cite{kontsevich,PoZa} of these
$\tht$-functions is that they count the areas of the disk
instantons that are bounded by the three intersecting $D$-branes
$\cL_i$ ($q=e^{2\pi i\rho}\sim e^{-2\pi\,{\rm Area}}$).
This is visualized in Fig.\ \ref{triangles}. The $\xi$-dependence 
takes position shifts and Wilson lines on the $A$-branes into account.  
The expressions \eqref{cijk} coincide with the Yukawa couplings given in 
\cite{CIM}, which were obtained by a direct evaluation of the areas of 
the triangles and~summing~them~up.

\goodbreak
\section{Fukaya products and $\tht$-identities}
\nobreak
One may wonder what the underlying mathematical reason is why the
triple matrix product of the $\Psi$'s yields the correct disk
partition functions (\ref{cijk}). We have already mentioned that
the result arises due to non-trival addition formulae of $\tht$-functions
on which the $\Psi$'s depend. On the other hand
 we know from \cite{PoZa,Poli,Poli1,Poli2,Poli3}
that certain such formulae represent Fukaya products of
the derived category on the elliptic curve.  It is thus desirable
to exhibit this connection more explicitly, by identifying the
kind of $\tht$-function identities that underly our results.

Specifically, for general vector bundles on the elliptic curve, the
first non-zero, associative Fukaya product $m_2:\
$Hom$[\cL_i,\cL_j]\otimes$Hom$[\cL_j,\cL_k]\rightarrow$Hom$[\cL_i,\cL_k]$
can be written in the following form \cite{PoZa,Poli,Poli1,Poli2}:\footnote{In 
this section, we will denote the K\"ahler parameter on the
$A$-model side by $\tilde\rho=\tau$, where $\tau$ is the complex
structure parameter in the $B$-model.}

\beql{m2prod}
m_2([e_{ij}(0,a),[e_{jk}(0,b)])\ =\ \sum_{n\in I_{\la_j}/I_{\la_i\la_j\la_k}}
\tht_{I_{\la_i\la_j\la_k};n}(p\tilde\rho)\,[e_{ik}(n,-\la_jn+a+b)]\ ,
\eeql
where $[e_{ij}(m,k)]$ denote basis elements of Hom$[\cL_i,\cL_j]$,
and the arguments denote certain lattice 
shifts further explained in \cite{Poli,Poli1,Poli2}.
Moreover, $\la=c_1/r$ denotes the slopes of the branes, and
$I_{\la_i}=\{n\in\ZZ: n\la_i\in\ZZ\}$,
$I_{\la_i\la_j\la_k}=I_{\la_j}\cap\frac{\la_k-\la_i}{\la_k-\la_j}\la_i$.
Furthermore, $p=\frac{(\la_k-\la_j)(\la_j-\la_i)}{\la_k-\la_i}$ and
$\tht_{I;n}$ denotes a $\tht$-function of the form (\ref{THETAdef}), but for
which the sum runs over $m\in I+n$. 

The product (\ref{m2prod}) takes the form of a $\tht$-function
identity when the basis elements $[e_{ij}]$ are represented by
sections made out of $\tht$-functions. This is particularly simple
for line bundles, where $r(\cL_i)=1$, $\la_i\in\ZZ$ and
for which the $[e_{ij}]$ are directly given by $\tht$-functions: $[e_{ij}(0,a)]\sim
\tht_{(\la_j-\la_i)\ZZ,a}(\frac{\tilde\rho}{\la_j-\la_i})$. For more general
vector bundles with $r(\cL_i)>1$ (which applies to our example), 
one needs to employ isogenies (rescalings of $\tilde\rho$),
and consider $r$-tuples of sections; see \cite{PoZa,Poli,Poli1,Poli2,Poli3}
for details.

For the case at hand, we identify the labels as $(i,j,k)=(2,1,3)$, and
we have for the slopes $\la_\ell=\lambda(\cL_\ell)$
of the bundles: $\la_1=1/2$, $\la_2=-1$, $\la_3=2$ which yields
$p=3/4$.  Because of $I_{\la_\ell}=2\ZZ=I_{\la_2\la_1\la_3}$, the sum
in (\ref{m2prod}) runs only over $n=0$. The resulting $\tht$-function
on the RHS of (\ref{m2prod}), given by $\tht_{2\ZZ,0}(3/4\tilde\rho)$,
precisely coincides with the Yukawa couplings given in the previous
section.  Moreover, the $[e_{ij}]$ can be represented by sections
of Hom$[\cL_i,\cL_j]\cong H^0(L^{\otimes3})$, which is three-dimensional
and is generated by $\al_\ell(-u_j-u_i)$.

To make contact with our Landau-Ginzburg computations, notice that the result
(\ref{cijk}) for the correlators (\ref{threepoint}) can be expressed
by the following operator product:
\beql{matrixp}
\Psi_{21}^{(a)}(u_2,u_1)\cdot\Psi_{13}^{(b)}(u_1,u_3)\ =\ 
\sum C_{ab}^{\ \ c}(\tilde\rho,u_1+u_2+u_3)\,\Phi_{23}^{(c)}(u_2,u_3)\ ,
\eeql
(modulo $Q$-exact pieces).  This is nothing but the $B$-model mirror
Landau-Ginzburg representation of the Fukaya product (\ref{m2prod}).
The $[e_{ij}]$ are represented here by the matrix-valued, $\ZZ_3$
equivariant sections $\Psi_{ji}^{(a)}$ and $\Phi_{ij}^{(a)}$ as
given in Section~2, with the proper normalizations.  Specifically,
recalling that the $\Psi$'s were already rescaled by $c_{ij}$ defined
below (\ref{zetaident}), and implementing the rescaling mentioned
at the end of the previous section, it follows that the normalization
functions in $\cQ$ for the fermionic ring elements can be chosen
as follows:
\beql{fernorm}
g^{(b)}_{i+1\,{\rm mod}\,3,i}\ = \ \Big(
\frac {q(\tilde\rho)}{\eta(\tilde\rho)}\Big)^{-1/3}\ 
=\ 
{\rm const.}(\bpsi')^{1/4}(1-\bpsi^3)^{-5/24}\ ,\ \ b=1,2,3,
\eeql
where we have used $\eta^8=\frac3{(2\pi i)^2}\frac{(\bpsi')^2}{\bpsi^3-1}$.
The bosonic normalizations are then fixed by (\ref{topmetric}), and
in particular we find:
\beql{bosnorm}
g^{(1)}_{23}\ =\ {\rm const.}(\bpsi')^{1/4}(1-\bpsi^3)^{-7/24}
\Big(\al_2(u_2)\al_2(u_3)\al_3(u_2)\al_3(u_3)\al_3(-u_2-u_3)\Big)^{-1}.
\eeql
Using these normalizations, and by repeatedly using the cubic
equation (\ref{alpha}), we find that, for example, the product
$m_2(\Psi^{1}_{21}\Psi^{1}_{13})=\al_1(u_1+u_2+u_3)\Phi^{1}_{23}$
boils down to the following identity between $\tht$-functions:
\beql{thetaid}
&&\frac1{\eta(\tilde\rho)\al_1(u_1)}
\Bigg(
\frac{\al_2(-u_3-u_1)\al_3(-u_2-u_1)}{\al_2(u_2)\al_3(u_3)}
-
\frac{\al_2(-u_2-u_1)\al_3(-u_3-u_1)}{\al_2(u_3)\al_3(u_2)}
\Bigg)\\ &&\qquad= \qquad
\al_1(u_1+u_2+u_3)\,
\frac {\al_1(u_3-u_2)}{\al_2(u_2)\al_2(u_3)\al_3(u_2)\al_3(u_3)}\ ,
\eeql
whose left- and right-hand sides correspond to eq.~(\ref{m2prod});
the other products lead to analogous expressions, and we do not
need to present them here. Noting that the denominator on the RHS
stems from the normalization of $\Phi^{1}_{23}$, we see that the
structure of the RHS is quite simple; this is a reflection
of the fact that the involved bundles $\cL_{2}\sim \cO(-1)$,
$\cL_{3}\sim \cO(2)$ are line bundles, for which the morphisms are simple
$\tht$-functions. On the other hand, the LHS ist structurally more
complicated, and this reflects the involvement of $\cL_1$
which is a rank two bundle.\footnote{Note that the
components of the $\Psi_{ij}^{(a)}$ (after rescaling (\ref{zetaident}))
depend only on $\al_\ell(-u_i-u_j)$ and
$\frac{\al_\ell(-u_i-u_j)}{\al_k(u_i)\al_m(u_j)}$. The latter
expression may be viewed as a higher-degree version of the Kronecker
function, and indeed it is known that $\theta$- and Kronecker
functions are the natural sections of rank two bundles on the
elliptic curve \cite{Poli3}.}

Summarizing, we have demonstrated that the boundary Landau-Ginzburg
approach reproduces non-trivial mathematical results about the
category of $D$-branes on the elliptic curve. We expect it to capture
branes with higher $K$-charges and also the higher products $m_k$
(though likely with considerably more effort),
 as well as generalizations to branes
on higher dimensional manifolds; this will be discussed elsewhere.


\begin{acknowledgments}
We would like to thank Kentaro Hori and Calin Lazaroiu for collaborative
input into this project. We are most grateful to Alexander Polishchuk and 
Daniel Roggenkamp for enlightening discussions. I.B.\ thanks ETH Z\"urich 
for hospitality, and W.L.\ and J.W.\ acknowledge the ICMS in Edinburgh and
the organizers of the very stimulating Workshop on Derived Categories, 
Quivers and Strings at which parts of this work were done.
This research was supported in part by the National 
Science Foundation under Grant No.\ PHY99-07949.
\end{acknowledgments}

\goodbreak
\appendix

\vskip 2em
\nobreak
\appendix
\noindent
{\bf\large Appendix: LG description of the short diagonals.}
\vskip 1em

The $3\times 3$ matrix factorizations discussed in the main part
of the paper do not describe the minimal branes, \ie, the generators
of the full $K$-charge lattice on the torus. These have slopes
$(r,c_1)=(1,0)$ and $(0,1)$, corresponding to pure $D2$ and $D0$
branes, and are the $B$-model mirrors of the short diagonals $\cS_1$,
$\cS_2$, $\cS_3$ of the $SU(3)$ torus, as shown in Fig.~1. The
minimal branes do not arise as pull-backs from the ambient $\IP^2$,
but are intrinsically tied to the curve $W=0$ in $\IP^2$.

In this appendix, we study a class of $\zet_3$-equivariant,
quasi-homogeneous $2\times 2$ matrix factorizations of the cubic 
\eqref{cubicdef}, that describe these minimal branes.
At $a=0$, such matrix factorizations were
discussed in \cite{laza}. Analogous branes for the quintic
at the Fermat point have been obtained in \cite{dAAD}, where it
was shown that they provide an integral basis of the full
charge lattice. 
 
We start with the following system of
homogeneous linear functions:
\begin{eqnarray*}
L_1 &=& \alpha_3 x_1 - \alpha_2 x_3 \\
L_2 &=& -\alpha_3 x_2 + \alpha_1 x_3.
\end{eqnarray*}
For $\alpha_3=0$ the linear equations $L_1=L_2=0$
describe a point which lies on the torus provided that the $\alpha_i$
fulfill the torus equation (\ref{cubicdef}).
We can then find two polynomials $F_1,F_2$
of degree $2$ such that
$$
\alpha_1 \alpha_2 \alpha_3 W= L_1 F_1 +L_2 F_2.
$$
Explicitly, $F_1, F_2 $ can be chosen to be
\begin{eqnarray*}
F_1 &=& \alpha_1 \alpha_2 x_1^2 + \alpha_2^2 x_1 x_2 - \alpha_1^2 x_2^2
-\alpha_1 \alpha_3 x_3^2 \\
F_2 &=& \alpha_2^2 x_1^2 - \alpha_1^2 x_1 x_2 - \alpha_1 \alpha_2 x_2^2
+\alpha_3^2 x_1 x_3.
\end{eqnarray*}
Under the exchange $x_i \leftrightarrow \alpha_i$ the polynomials transform
as $L_1 \leftrightarrow L_2$ and $F_1 \leftrightarrow -F_2$.

Note that the factorization becomes singular in the limit $\alpha_3\to
0$, since the equations $L_1=L_2=0$ fail to describe a point in
that case.  To cover this coordinate patch, one has to use linear
combinations of $L_1, L_2$ that are well-behaved in the limit, such
as the system consisting of $\tilde L_1$ and $\tilde
L_2=\frac{1}{\alpha_3}(\alpha_1 L_1+\alpha_2 L_2)$ and $\tilde F_1=F_1-\frac{\al_1}{\al_2}F_2$
and $\tilde F_2=\frac{\al_3}{\al_2}F_2$.

The BRST operator takes the form
$$
Q= \tilde L_1 \pi _1 + \tilde L_2 \pi_2 +  
\frac{1}{\alpha_1\alpha_2\alpha_3}(\tilde F_1\bar\pi_1 + \tilde F_2 \bar\pi_2),
$$
where $\pi_i,\bar\pi_i$ form a representation of the four dimensional Clifford
algebra. It can also be written in the form (\ref{simpleQ}) with 
$J=\left(\begin{smallmatrix} \tilde L_1 & \tilde F_2 \\ 
-\tilde L_2 & \tilde F_1 \end{smallmatrix}\right)$
and
$E=\frac{1}{\alpha_1\alpha_2\alpha_3}
\left(\begin{smallmatrix}\tilde F_1 & -\tilde F_2 \\
\tilde L_2 & \tilde L_1\end{smallmatrix}\right)$.

To verify that the $2\times 2$ factorizations correspond to
the short diagonals in the $A$-picture, 
we determine their charges. This can
easily be done by first determining their intersection numbers with the
$3\times 3$-factorization type of branes. In a second step one can
then determine a collection of $3 \times 3$ branes 
which have the same intersection numbers with any other
set of $3\times 3$ branes as the $2\times 2$ branes.   
The charge of this collection of $3\times 3$ branes is known
from our earlier considerations and equals the charge
of the $2\times 2$ branes.

The intersection numbers of the $2\times 2$ factorizations
with the $3\times 3$ factorizations have been determined in
\cite{dAAD}. In that paper, all computations were done exclusively
at the Gepner point, but since the intersection numbers are
topological, we can make use of their calculations. The result is
that the intersection matrix is
\begin{equation}\label{intersect}
I_{2\times 2, 3\times 3} = -1 + 2g -g^2,
\end{equation}
where $g$ is the $\ZZ_3$ shift matrix that shifts the $\ZZ_3$ representation
label of a brane by one. For our calculation, we need in addition
the intersection matrix of the $3\times 3$ branes, which is given
by
$$
I_{3\times 3, 3\times 3} = -3g+3g^2.
$$
We now look for a stack of $x_i$ branes of type $\cL_i$ having th
intersection numbers (\ref{intersect}). This amounts to the following equation:
$$
(-3g+3g^2)(x_1 + x_2 g + x_3 g^2) = -1 + 2g - g^2,
$$
with the solution $x_1= -\frac{2}{3} + x_3, x_2= -\frac{1}{3} + x_3$.
Translating this into charges, the first of the three $2\times 2$ branes
has the charge of $q_1=-1/3(2q(\cL_3) +q(\cL_1))=(0,1)$ and is a pure
D0 brane, confirming the expectation that one of the branes should be
a pure D0 brane. The charges of the other two branes are
$q_2=-1/3 (2q(\cL_2) + q(\cL_3))= (1,0)$, which is a pure D2 brane,
and $q_3=-1/3(2q(\cL_1)+ q(\cL_2))= (-1,-1)$. To find the interpretation
of the branes in the A-type picture, we note that $2$ times a
long diagonal plus $1$ times the $\ZZ_3$ rotated long diagonal yields 
$3$ times a short long diagonal, such that the $2\times 2$
factorizations indeed 
correspond to the branes $\cS_i$ wrapped along the short diagonals (see Figure 1).

We can give another consistency check of our results by determining the
flat brane modulus as we did in section \ref{flat} for the $3\times 3$
factorizations. In the same notation, and in the same normalization as 
in eq.\ \eqref{disk}, we find for the $2\times 2$ factorizations
\begin{equation}
\langle\del_u Q\rangle_{\rm disk, normalized}= \frac 13 f(\tau,u)\,,
\end{equation}
where $f$ is as in \eqref{result}. The $\alpha_\ell$, then, have to be 
identified with $\tht$-functions as in \eqref{alphaTHETA}, with $3u\to u$.


\begingroup\raggedright\endgroup

\end{document}